
\overfullrule = 0pt
\catcode`\@=11
\expandafter\ifx\csname @iasmacros\endcsname\relax
        \global\let\@iasmacros=\par
\else   \immediate\write16{}
        \immediate\write16{Warning:}    \immediate\write16{You have
tried to input iasmacros more than once.}       \immediate\write16{}
\endinput
\fi
\catcode`\@=12




\def\Rf{\normalbaselines\parindent=0pt \medskip\hangindent=3pc \hangafter=1 }

\def\nonarrower{\advance\leftskip by-\parindent
        \advance\rightskip by-\parindent}


\def\boxit#1{\vbox{\hrule\hbox{\vrule\kern3pt
        \vbox{\kern3pt#1\kern3pt}\kern3pt\vrule}\hrule}}

\def\hence{\leavevmode\hbox{\bf .\raise5.5pt\hbox{.}.} }
\def\solar{\odot}

\def\dalemb#1#2{{\vbox{\hrule height.#2pt
        \hbox{\vrule width.#2pt height#1pt \kern#1pt \vrule
width.#2pt}     \hrule height.#2pt}}}
\def\gtorder{\mathrel{\raise.3ex\hbox{$>$}\mkern-14mu
             \lower0.6ex\hbox{$\sim$}}}
\def\ltorder{\mathrel{\raise.3ex\hbox{$<$}\mkern-14mu
             \lower0.6ex\hbox{$\sim$}}}

\newdimen\fullhsize
\newbox\leftcolumn
\def\twoup{\hoffset=-.5in \voffset=-.25in
  \hsize=4.75in \fullhsize=10in \vsize=6.9in \def\fullline{\hbox
to\fullhsize} \let\lr=L \output={\if L\lr
\global\setbox\leftcolumn=\columnbox\global\let\lr=R \advancepageno
\else \doubleformat \global\let\lr=L\fi \ifnum\outputpenalty>-20000
\else\dosupereject\fi} \def\doubleformat{\shipout\vbox{
\fullline{\box\leftcolumn\hfil\columnbox}\advancepageno}}
\def\columnbox{\leftline{\vbox{\makeheadline\pagebody\makefootline}}}
\tolerance=1000 }

\baselineskip 12pt
\def\et{{\it et al.~}}
\def\page{\vfill\eject}
\def\msun {M_{\odot}}
\def\DM {{_{DM}}}
\def\MO {{_{MO}}}

\narrower
\centerline{ }
\vskip 0.7in
\centerline{\bf Can the Dark Matter be $10^6$ Solar Mass Objects?}
\vskip 0.6in
\centerline{\it Hans-Walter Rix\footnote{$^1$}{\rm Hubble Fellow}}
\smallskip
\centerline{\it Institute for Advanced Study}
\smallskip
\centerline{\it Princeton, NJ, 08540}
\bigskip
\centerline{\it and}
\bigskip
\centerline{\it George Lake}
\smallskip
\centerline{\it Dept. of Physics, University of Washington}
\smallskip
\centerline{\it Seattle WA, 98195}
\vskip 0.3truein
\centerline{\bf Abstract}
\smallskip
If the dark matter in galactic halos is made up of compact,
macroscopic objects (MO), such as black holes
with $M_\MO >>M_{stars}$, gravitational
scattering will lead to kinematic heating of the stars.
Observational constraints on the amount of heating in the disk
of the Milky Way put
upper limits on $M_\MO \ltorder 10^{6.3}\msun $.
We find limits that are three orders of
magnitude more stringent by examining the heating limits in low mass
stellar systems, where higher densities of dark matter
and lower relative velocities
would destroy stellar disks or disperse the stars in less than a billion years.

Limits on $M_\MO$ are derived from two nearby dwarf galaxies,
dominated by dark matter: the presence of a flat stellar disk in the dwarf
spiral galaxy DDO~154 is shown to imply $M_\MO \ltorder 7\times
10^{5}\msun$, comparable to the limits derived for the Galactic disk.
However, the structure and kinematics of the Local Group member GR8
yield a limit of $M_\MO \ltorder 6\times 10^{3}\msun$.

We also examine the possibility that the local disk heating is done by compact
clusters of brown dwarfs rather than black holes.  Such clusters could
dissolve in the higher density halos of small galaxies.
While theoretical arguments have been presented for such clusters,
they should have been
detected in the IRAS point source catalog.

If the properties of the dark matter are universal these results preclude the
dominance of dark matter constituents in the cosmologically interesting mass
range $\sim 10^6\msun$ and limit them to $M_\MO \ltorder
10^{3.7}\msun$. These results also rule out massive compact halo
objects as significant contributors to the kinematic heating of the
Galactic disk.

\bigskip
\centerline{\bf I. Introduction}
\smallskip
The self-gravity of many astrophysical systems is clearly dominated by
dark matter.
Starting from Zwicky's original claim of dark matter in clusters of galaxies,
evidence for dark matter has been found on both larger scales and on smaller,
galactic scales (see e.g. Trimble 1987 and Ashman 1992 for reviews).

While the existence of dark matter has been clearly established,
its nature is still unknown.
Candidates currently under discussion range from postulated
elementary particles to compact objects with masses vastly greater
than stellar masses--a mass range of $10^{70}$!
Compelling cases have been made for candidates throughout this
mass range.  Our current effort is directed at constraining this
vast range of possible masses.

Here we examine the case for
``macroscopic"
dark matter candidates, ranging from compact objects with substellar
masses to objects with masses in
excess of $10^6\msun$, possibly massive black holes. Small objects and
black holes are particularly attractive candidates, because they have a low
luminosity per unit mass and are thus ``dark".  The existence and
mass-scale of such compact massive objects can be constrained
principally by two observable effects: gravitational lensing and
gravitational relaxation.

\noindent Efforts to detect these effects have mostly focussed on the halo of
the Milky Way:

\noindent$\bullet$ a search for gravitational lensing amplification of
background stars from compact halo objects in the mass range from
$10^{-6}\msun$ to $10^{2}\msun$ is currently being attempted by
groups (see e.g. Alcock \et 1993).  An extension to larger masses has
been considered by Gould (1992).

\noindent$\bullet$ black holes with masses of $2\times 10^6\msun$
would provide a source for the gravitational heating needed to
explain the increase in velocity dispersion with age of the stellar
disk population (Lacey \& Ostriker 1985).
Insisting that they do not heat the disk too quickly places
an upper limit on the typical mass of compact halo
objects of $10^{6.3}\msun$.

A variation on this scenario has been
proposed by Carr and Lacey (1987), who suggested compact clusters of
dark objects instead of single black holes. They advocated such
clusters because they can dissolve
before they collect at the galactic center due to dynamical friction
(see also Hut and Rees, 1993) and
because gas accretion events during the passage of such objects through
a molecular cloud in the Milky Way are less luminous in X-rays.
As far as gravitational relaxation is concerned, tightly bound
clusters of macroscopic dark matter are indistinguishable from point masses
and all results derived below apply equally to both cases.

At present there are no firm dynamical limits on compact objects
with $M_\MO \ltorder10^{6.3}\msun$ comprising the ubiquitous dark matter.
However, mass-scales near this upper limit are
particularly interesting, since they correspond to the Jeans mass just
after the epoch of recombination and thus may collapse to great
overdensities at high redshifts (e.g. Dicke and Peebles 1968, White
and Rees 1978, Ashman and Carr 1988).

In this paper we show that progress can be made in
limiting $M_\MO$ by
considering smaller and less massive stellar systems than the Milky Way
that nevertheless contain large amounts of dark matter.  There,
relaxation due to two-body encounters between the dark compact objects
and the stars is stronger for two reasons: first, for a given $M_\MO$
less massive galaxies contain a smaller number
of such massive objects and hence have a ``grainier" potential well.
Second, small systems have lower velocity dispersions
making the energy exchange in two-body encounters more efficient.

The arguments presented here hold under two assumptions:
(1) the properties of the dark matter are universal and (2)
a characteristic mass scale, $M_\MO$, can be assigned to the dark matter
constituents.
The kinetic ``heating rate" due to
objects of a given mass is proportional to their (individual) mass
and their contribution to the mass density (Eq. 1). If there is a range
of masses, say, with number density $N\propto M^\gamma$ for
$M_{light}<M_\MO<M_{heavy}$,
the light objects will
dominate the heating if $\gamma<-3$, the heavy ones otherwise.

\vskip 0.15truein
\centerline{\bf II. Relaxational Heating}
\vskip 0.05truein
\centerline{\it II.1. Basics}

In this section we will review a few relations describing the
statistical effects of two-body encounters in stellar dynamical
systems.  For a detailed discussion we refer to Chandrasekhar (1960),
Lacey and Ostriker (1985) and Spitzer (1987). The basic assumptions
made for the present work are that $M_\MO \gg M_{stars}\sim \msun$
and that the dark matter is assumed to dominate the mass density
and hence the stars can be viewed as test particles.
The net effect of these statistical encounters
will result in an approach to kinetic equipartition, increasing the
{\it rms} velocity of the observable stars by up to $\sqrt{M_\MO/M_{stars}}$.

We also assume that the dark matter velocity distribution is approximately
Maxwellian and has a dispersion substantially larger than the stars'.
Then the expected one-dimensional energy increase
(per unit mass) of the stars can be found as (see e.g. Spitzer 1987, Eq. 2-60):
$${\langle \dot E_S\rangle} = \sqrt{32\pi} ln\Lambda~ G^2 M_\MO
\rho_\DM \sigma^{-1}_\DM
\eqno(1)$$
where $ln\Lambda$ is the Coulomb logarithm, $M_\MO $ is the mass of
the individual dark matter constituents, $\rho_\DM $ is their mass density, and
$\sigma_\DM$ is the one-dimensional
velocity dispersion of the dark matter.
For relaxation of a system of N equal mass particles the Coulomb logarithm is
$ln\Lambda\approx ln{\rm N}\approx ln(M_{tot}/M_\MO)$ (e.g. Binney and Tremaine
1987),
where $M_{tot}$ where is the total mass of dark matter in the volume under
consideration. As detailed in Section III, we will deal with systems
of $M_{tot}\sim 10^7-10^8 \msun$ and will be interested in objects with
$M_\MO\sim 10^4-10^5\msun$, implying $ln\Lambda\approx 7$.
This value is quite similar to the value of 10 used by Lacey and Ostriker
(1985).

A limit on $M_\MO $ can then be obtained by demanding that: $$
\int{\langle \dot E_S\rangle}dt < \sigma^2_{S}/2, \eqno(2)$$
where $\sigma^2_{S}$ is the velocity distribution of the stars
and the integral extends over the lifetime of the stars.

\vskip 0.5truein
\centerline {\it II.2. Halo Model}

We assume that the observable stars reside within the core of a halo
with a mass profile: $$\rho (r) = { \rho_\DM \over \Bigl[{1+({r\over
a})^2}\Bigr]^2 } \eqno(3)$$ (e.g. Richstone and Tremaine 1986). This
model has core radius of $R_\DM =0.77a$, a total
mass of $M_{tot} = \pi^2\rho_\DM a^3$ and a central velocity
dispersion of $$\sigma_\DM^2 = 1.6G\rho_\DM R^2_\DM.\eqno(4)$$ For a
given luminosity L of the whole system, the central velocity dispersion
scales as $$\sigma_\DM\propto \rho_\DM^{5\over 6}\cdot (M/L)^{1\over
3}.\eqno(5)$$

\smallskip
\noindent The observational input needed in Eqs. 1 and 2 consists of
(a) the kinetic energy of the tracer stars, $\sigma_S$, and (b)
the central mass density of the halo, $\rho_\DM$, or
a lower limit on it.  Test particles (stars, gas) with an isothermal
distribution function of dispersion $\sigma_S$ residing in the dark matter core
will have a Gaussian radial profile with width $H$.  The central mass
density of the halo is then given by $$ \rho_\DM={
{3\sigma_S^2}\over{4\pi G H^2} }.\eqno(6)$$

\noindent$\bullet$ The velocity dispersion of the halo
constituents, $\sigma_\DM$, or an upper limit on them. According to
Eqs. 4 and 5 this is best phrased in terms of the total $M/L$ of the
system.

Rewriting Eq. 1 in terms of the observable quantities, the increase
in velocity dispersion of the kinematic tracers (e.g. stars)
after time $\Delta t$ is
$$\eqalign{
\Delta \sigma_S^2 = (28~{\rm km/s})^2\cdot &\bigl({{\rho_\DM}\over{5\times
10^{-2
   4}g/cm^3}}\bigr)^{5\over 6}
\bigl({{M_\MO}\over{10^5M_\solar}}\bigr)\cr
&\bigl({ {M/L}\over{130M_\solar/L_\solar} }\bigr)^{-{1\over 3}}
\bigl({ {L}\over{2.5\times 10^5/L_\solar} }\bigr)^{{1\over 3}}
\bigl({ {\Delta t}\over{10^9yrs}}\bigr).}\eqno(7)$$

The particular value for parameterizing M/L is taken from Lake
(1991). The M/L should, however, be viewed as an essentially free
parameter within an order of magnitude.
It is worth noting how these limits depend on the adopted distance to
the galaxy. Equation 7 implies that
the derived limit on $M_\MO$ scales as D$^2$ through the
observables $\rho_\DM$ and $L$.

\vskip 0.20truein
\centerline{\bf III. Small Stellar Systems Dominated by Dark Matter}
\vskip 0.1truein

\bigskip
\centerline{\it III.1 The Dwarf Spiral Galaxy DDO~154}

Carignan and Beaulieu (e.g. 1989, hereafter
CB89) and Lake \et (e.g. 1990) have drawn attention to the dynamics of
a class of small gas-rich spiral galaxies (e.g. DDO154, DDO170) whose
HI kinematics indicate the presence of large amounts of dark matter.
At an adopted distance of $4$~Mpc, DDO~154 has a
luminosity of $L_B\sim 5\times 10^7L_\solar$ and $\rho_\DM\sim
0.009\msun /pc^3$; however, it must be noted that the distance to
DDO154 is uncertain by a factor of about two.

Since stellar disks are kinematically ``coldest" in the $z$
direction, the most drastic effect of relaxational heating will be the
increase in the velocity dispersion perpendicular to the plane.
The stellar surface brightness in DDO~154
is too low for kinematic measurements. However, an increase of the
velocity dispersion will lead to an increased vertical scale height of
the stellar distribution.  We can use the intrinsic flattening of
the stellar distribution to put upper limits on their z velocity
dispersion. From fitting the HI rotation curve CB89 derive an
inclination of the system of $63^{\circ}$ is derived, while the axis
ratio of the light at two disk scale lengths is 0.57. This implies
an intrinsic thickness  of $\sim H_z/R_{exp}=0.3$, where $H_z$ denotes the
vertical
scale height and $R_{exp}$ is the radial scalelength.
The vertical velocity dispersion
at a radial scale length is given by
$\sigma_S\sim v_{circ}\times H_z/R_{exp}$, if the self-gravity of the
disk is neglected.  This yields $\sigma_S\sim 17km/s$, for a circular
velocity of $\sim 50km/s$.

The age of the stellar disk may be constrained by its azimuthal
smoothness.  I($0.8\mu$) band images (Rix, {\it unpublished}) show
that $\Delta I/I<0.1$ along an elliptical annulus.  Consequently,
the age of the stellar disk must be at least of order
five orbital periods to allow for the azimuthal smoothing of
the disk. At 2kpc the orbital time is $\sim 3\times 10^{8}yrs$ and so we
adopt $\Delta t \gtorder 1.5$Gyr.

Using the above parameters and Eq.~7, the present flatness of the observed
stellar
disk implies for the mass of the individual dark matter constituents: $$M_\MO
\ltorder 7\times 10^5\msun~\Bigr[{{\rho_\DM }\over{0.009\msun
/pc^3}}\Bigl]^{-5/6} \times \Bigr[{{1.5\times 10^9yrs}\over{\Delta
t}}\Bigl] \times \Bigr[{{\Delta\sigma_z}\over{17{\rm
km/s}}}\Bigl]^{2}.\eqno(8)$$

We have used here a parameterization different from Eq. 7,
because $\sigma_S$ is
inferred indirectly.  Note that the dependence of $M_\MO$ on the
adopted distance enters only linearly in this estimate if the lower
limit on the disk age is derived from kinematic considerations.
Similar limits can be derived for DDO~170 (Lake \et 1990).

Even though DDO~154 is a much lower mass object than the
Milky Way, the limits on $M_\MO$ are not much more stringent than the ones
derived from the Galactic disk. This is because we cannot place
stringent limits on both $\sigma_S$ and on the age of the disk
population.

\bigskip
\centerline{\it III.2 The Gas Rich Dwarf Galaxy GR8}

By far the best system in the local group to derive limits on $M_\MO$ is GR8.
Its distance is estimated as 1.1Mpc (de~Vaucouleurs and Moss, 1983)
with an uncertainty of $\sim 50\%$,
(see also Hoessel and Danielson 1983, who advocate a distance of
1.4Mpc). For a distance of 1.1Mpc it has a
luminosity of $L_B \sim 2.5\times 10^6L_\solar$.  It is the most
extreme, gas-rich dwarf mapped in HI (Carignan, Beaulieu and Freeman
1990).
In the inner parts ($R<250$pc) the HI gas rotates at a speed of
$6-8$km/s, becoming increasingly pressure supported with
$\sigma_{HI}\sim 10$km/s throughout (The rotation drops
below $3km/s$ at 500pc).
Its radial surface density profile can
be fit by a Gaussian with $H=294$pc.  The stars have an exponential
light profile with $R_{exp}(B)=76pc$.

Although the velocities and dispersions measured for the HI gas are
small ($\ltorder 10km/s$, comparable to the turbulent velocities
in Galactic HI), they can nevertheless be used for dynamical mass estimates.
In our Galaxy, HI clouds need not be in virial equilibrium
because they can be confined by the hot phase of the ISM, which
in turn is confined by the deep potential well of the Galactic
halo. Since, GR~8 is much too small a galaxy to retain a hot, gaseous halo, the
HI in it cannot be pressure confined and must hence be held together by
gravity.

Carignan, Beaulieu and Freeman (1990) used the Jeans
equation for radial equilibrium to account for the radial
surface density profile of the HI as well as for its rotation and
dispersion and inferred a mass profile. The dynamical mass is found to
exceed greatly the observed stellar
and HI mass, indicating the presence of dark halo.
Using Eq. 6, one finds for the central dark matter density
$\rho_\DM\sim 5\times 10^{-24}gcm^{-3} \sim 0.07M_\solar pc^{-3}$
(Carignan, Beaulieu and Freeman 1990).  The kinetic energy of the
stars can be estimate through
$\sigma_S=\sigma_{HI}\times\bigr(R_{0.5}(T)/R_{0.5}(HI)\bigl)=3.7$km/s,
where $R_{0.5}$ denotes the half-light, or half-mass, radius of the
stars and the neutral gas.

Requiring that $\Delta\sigma^2<\sigma_S^2$, Eq. 7 yields
$$M_\MO\ltorder6\times 10^3M_\solar \Bigl({{10^9yrs}\over{\Delta
t}}\Bigr)
\Bigl({{M/L}\over{130M_\solar /L_\solar}}\Bigr )^{1\over 3}\eqno(9)$$
as the limit on the masses of the halo constituents.

Theoretical arguments and observational constraints can be used to place
a lower limit on the age
of the stellar population in GR8.
Lake (1991) has shown that the
redshift of the expected turnaround epoch for a system of GR8's
central density and of ${\rm M/L}\sim 130$ is $z_{turn}>7$. Even for
extreme assumptions about M/L, one expects $z_{turn}{\rm (GR8)}\gtorder
z_{turn}{\rm (MW)}$. At the epoch of turnaround the gas in GR8 had a
higher density and could cool more efficiently that the Milky Way's gas.
As a result, the galaxy could collapse and cool to its current radius much
earlier than the Milky Way.  Star formation should occur promptly if
the galaxy obeys the empirical star formation laws observed for larger
spirals (Kennicutt 1989).  We note that Skillman (1993) finds that these
empirical laws are supported by the star formation in the extreme dwarf
galaxy IC 1613.
Hence, it is most likely that star formation in GR8 started at least
as early as in the Milky Way disk.

This scenario is supported empirically by two observations:
In their color-magnitude diagram of GR~8 Hoessel and
Danielson (1983) find evidence for the presence of carbon stars: a
group of red stars with V=22.0mag and B-V=2.0, just as expected for
carbon stars at GR8's distance. Since carbon stars are evolved stages of low
mass progenitors, they must be old.
Second, spectral population synthesis by Hunter and Gallagher (1985)
show that at $5000$\AA ~40\% of the light comes from O/B
stars and 60\% arises from G/K/M stars. They argue on this basis that
even though GR8 has had a very recent (and ongoing) episode of star formation,
it also contains a very significant old stellar population.

In this light it is most likely that ${\Delta t}>1$Gyr,
and consequently $M_\MO\ltorder6\times 10^3M_\solar$.  This
limit does not change much if we allow for the uncertainties in
the other parameters: even if we
considered extreme values for M/L, say 1000, and a distance
as large as 2Mpc, the derived limits are still two orders
of magnitude below what is
required for the local heating of the Galactic disk.

\bigskip
\centerline{\it III.3 Do ``dark clusters" provide a loophole?}

Rather than black holes, Carr and
Lacey (1987) proposed clusters having a total mass $10^{6.3}M_\solar$ and
typical size $\sim 1$pc, consisting of stellar or sub-stellar objects,
such as brown dwarfs.  Such clusters could circumvent our previous limits if
they
dissolved in these higher density dwarf galaxies.
For the particular values chosen by Carr and Lacey, clusters
will dissolve by mutual collisions in a Hubble time if the
galactic density is $\sim 1 \msun pc^{-3}$ (c.f. Hogan and Rees 1988).
If such clusters were larger by factors of a few, they could get disrupted
in GR8 ($\rho\sim0.1\msun pc^{-3}$), yet still survive at $R_0$ in the Galaxy.
By fine-tuning the cluster's density, one could conceive a scenario
whereby the clusters survive in the Milky Way's halo and dissolve in
the high density halos of dwarf galaxies.

Remarkably, we find that such clusters of brown dwarfs have been ruled out by
exploration of the IRAS point source catalog.
Beichman {\it{et al.}} (1990) have examined all sources that have faint
optical counterparts and found no brown dwarfs.
If the dark matter in the solar neighborhood were
brown dwarfs with
$\rho\sim 0.1 \msun pc^{-3}$ (Oort 1932, 1960; Bahcall 1984),
they find that, for a variety of brown dwarf mass functions,
there should have been between 0.12 and 2.4 objects in their sample.
If we instead adopt a dark matter density appropriate for the halo,
$10^{-2} \msun pc^{-3}$ (Caldwell and Ostriker 1981; Kuijken and
Gilmore 1989), the expected number falls by one order of magnitude.

If we put the brown dwarfs into clusters, the number of objects
detected in a magnitude limited sample increases by $N_{cl}^{1/2}$, where
$N_{cl}$
is the number of brown dwarfs in a cluster.  If the clusters had
a mass of $10^{6.3} \msun$, then there would be as many as $10^8$ objects
in each cluster {\it and the number that would appear in the IRAS
point source catalog would increase by $10^4$}.   Hence, we would have
expected that the IRAS catalog would have had between $10^2$-$10^4$
brown dwarf clusters if they were indeed responsible for the disk
heating. Clusters of planetary mass objects, $M<2M_{Jupiter}$, are still
permitted
by the IRAS data.

Nonetheless, we are led to conclude that clusters of brown dwarfs
are not a viable option for the dark matter in the Milky Way. The only
permissible
configurations are
clusters with $M_{cl}=10^{6.3}\msun$ (suitable for the local disk heating)
made of planets (to avoid IRAS detection) with a radius of $\sim 5$pc (to
dissolve in
GR8 but not at $R_0$ in the Milky Way.)

\bigskip
\centerline{\bf V. CONCLUSIONS}

The arguments presented above considerably strengthen the case that
the constituents of dark matter in galactic halos, if they are macroscopic,
must not have masses of more than a few $1000\msun$; these kinematic
limits are three orders of magnitudes lower than previous limits
derived from the heating of the Milky Way's disk. In particular, the
cosmologically interesting mass range for massive, dark halo
constituents, $M_\MO \sim 10^6\msun$, is inconsistent with the
dynamical state of the nearby dwarf galaxy GR8.

It was possible to improve these limits because the body of
information on the dynamics of low mass stellar systems, dominated by
dark matter, has improved dramatically over the last five years.  Seemingly
promising candidates, the
dwarf spiral galaxies DDO154 and DDO170 yielded
only $M_\MO \ltorder
7\times 10^5\msun$, comparable to the limits from the Milky Way disk.
However, the observations of Carignan \et (1990) of
the structure of the HI gas and the stars in GR8 imply $M_\MO <6\times
10^3\msun$. The data preclude the possibility that the entire dark matter in
GR8 is provided by a central object, but rather exist in an
extended distribution.

The mass range of $10^{-6}\msun < M_\MO < 10^4$ of potential massive
halo objects will be probed by microlensing experiments in the next
few years (Pazcynski 1986, Griest 1991, Gould 1992).
The presence of all possible dark matter constituents with $M_\MO
>10^{-6}\msun$, can therefore be tested.

These results also exclude massive compact halo objects as
significant contributors to the kinematic heating of the local
Galactic disk.  Finally, we found that the IRAS point source catalog
completely rules out large clusters of brown dwarfs.

\bigskip
\noindent{\bf Acknowledgments }

This paper has benefitted from discussions with Keith Ashman, Dan Maoz,
Andrew Gould and Piet Hut.  Support for this work was provided
by NASA through a Hubble fellowship (HF-1024.01-91A) from the Space
Telescope Science Institute, which is operated by AURA, Inc., under
NASA contract NAS5-26555.

\page
\centerline{\bf References }
\smallskip

\Rf{Ashman,~K.~M., 1992, PASP, 104, 1109.}

\Rf{Ashman,~K.~M. and Carr,~B, 1988, MNRAS, 234, 219}

\Rf{Alcock,~C. \et  1993, in {\it Robotic Telescopes for the 1990's} ed. A.V.
 Filippenko, ASP, San Francisco, p.193}

\Rf{Bahcall,~J.N. 1984, ApJ, 287, 926}

\Rf{Beichman,~C., Chester,~T., Gillett,~F.C., Low,~F.J.,
Matthews,~K.  and Neugebauer,~G., 1990, A.J., 99, 1569}

\Rf{Binney, J., \& Tremaine, S., 1987, ``Galactic Dynamics", Princeton Univ.
Press.}

\Rf{Caldwell,~J.R. and Ostriker,~J.P. 1981, Ap.J. 251, 61}

\Rf{Carignan,~C. and Beaulieu,~S., 1989, ApJ, 347, 760}

\Rf{Carignan,~C. and Beaulieu,~S. and Freeman,~K., 1990, AJ, 99, 178}

\Rf{Carr, B.~J., and Lacey,~C.~G., 1987, ApJ, 316, 23}

\Rf{Chandrasekhar,~S. 1960, {\it Principles  of Stellar Dynamics}, Univ. of
Chic
   ago,
Re-issued by Dover}

\Rf{de~Vaucouleurs,~G and Moss,~D. 1983, ApJ, 271, 123}

\Rf{Peebles,~P.~J.~E. and Dicke,~R. 1968, ApJ, 154, 891}

\Rf{Gnedin,~N. and Ostriker,~J., 1992, ApJ, 392, 442}

\Rf{Gould,~A. 1992, submitted to ApJ.}

\Rf{Griest,~K. 1991, Ap. J. 366, 421}

\Rf{Hoessel~J. and Danielson,~G. 1983, Ap. J., 271, 65}

\Rf{Hogan,~C.J. and Rees,~M.R. 1988, Phys.Lett.B, 205, 228}

\Rf{Hunter~D.~A. and Gallagher, J.~S.III. 1985, Ap.J.Supp., 58, 533}

\Rf{Hut~P. and Rees, M. 1993, MNRAS {\it in press}}

\Rf{Kennicutt,~R.~C. 1989, Ap.J., 344, 685}

\Rf{Kuijken,~C. and Gilmore,~G. 1989, MNRAS, 239, 571}

\Rf{Lacey,~C.~G. and Ostriker,~J.~P., 1985, ApJ, 299, 633}

\Rf{Lake,~G., 1990, MNRAS, 244, 701}

\Rf{Lake,~G., 1989, AJ, 98, 1253}

\Rf{Lake,~G., 1991, ApJL, 356, L43}

\Rf{Lake,~G., Schommer,~R. and van~Gorkom,~J., 1990, AJ, 99, 547}

\Rf{Oort,~J.H. 1932, Bull.Astron.Inst.Neth. 6, 249}

\Rf{Oort,~J.H. 1960, Bull.Astron.Inst.Neth. 15, 45}

\Rf{Pazcynski,~B. 1986, ApJ, 304, 1}

\Rf{Richstone,~D. \& Tremaine,~S., 1986, AJ, 92, 72}

\Rf{Skillman, E. D. 1993, {\it in preparation}}

\Rf{Spitzer,~L., 1987, {\it Dynamics of Globular Clusters}, Princ. Univ. Press}

\Rf{Trimble,~V., 1987, Ann.Rev.A.A., 25, 425}

\Rf{White,~S.D.M. and Rees,~M., 1983, MNRAS, 183, 341}

\vfill\eject\end